\documentstyle[12pt]{article}
\makeatletter
\let\chapter\hid@chapter
\makeatother
\input psfig.tex
\def\lsim{\lower.5ex\hbox{$\; \buildrel < \over \sim \;$}}
\def\gsim{\lower.5ex\hbox{$\; \buildrel > \over \sim \;$}}
\begin{document}
\pagenumbering{arabic}
\title{Spectral Softening due to Winds in Accretion Disks}

\author{Sandip K. Chakrabarti\\
S.N. Bose National Centre for Basic Sciences\\
JD-Block, Sector-III, Salt Lake, Calcutta 700091, INDIA}

\maketitle

\begin{abstract}

Accretion flows may produce profuse winds when they have
positive specific energy. Winds deplete matter from the
inner region of the disk and makes the inner region thinner,
optically. Since there are fewer electrons in this region, it becomes easier 
to Comptonize this part by the soft photons which are intercepted from 
the Keplerian disk farther out. We present a self-consistent 
picture of winds from an accretion disk and show how the spectra 
of the disk is softened due to the outflowing wind.

\end{abstract}

\medskip

\noindent Indian Journal of Physics, v. 72B (6), 565-569, 1998.

\section{Introduction}

Outflows are common in many astrophysical systems which 
contain black holes and neutron stars. Difference between stellar
outflows and outflows from these systems is that the outflows in these
systems have to form out of the inflowing material only, whereas
in stars outflows are `extensions' of the staller atmosphere.
Although a black hole does not have a hard surface, the centrifugal barrier
due to angular momentum behaves like one, and therefore mass 
loss associated with this barrier could be computed in the same way
as the mass loss from a stellar surface. For a detailed review
on accretion disks and associated outflow, see Chakrabarti [1,2].

Chakrabarti [3] classified  all possible solutions of a black hole
accretion and winds and showed that a large region of the
parameter space (spanned by specific energy ${\cal E}$ and angular momentum
$\lambda$) with positive ${\cal E}$, winds would form. Such 
solutions (both accretion and winds) have been amply verified by numerical simulations [4].
Winds have been observed by earlier numerical simulations from Keplerian 
accretion disks [5], although
generalized self-similar wind solution of K\"onigl [6] by Chakrabarti \& 
Bhaskaran [7] show that the outflows are more favourable
if the disk itself is sub-Keplerian. Chakrabarti \& D'Silva [8]
and D'Silva \& Chakrabarti [9] showed that magnetized flares 
close to the funnel wall in the accretion disk could produce
winds similar to that of the Sun.
This is because, energetically, it is equally easy or difficult for
an ordinary star or compact object to produce winds. Molteni,
Lanzafame \& Chakrabarti [10] showed through numerical 
simulation that up to about $15$ to $20$ percent of
mass loss is common in the case of weakly viscous, and thick accretion flows.
Chakrabarti [11] and Das \& Chakrabarti [12] gave a formulation
of the global inflow-outflow solutions [GIOS] and estimated the mass outflow 
rate as a function of the inflow parameters of the accretion flow. Particularly,
it was shown [12] that in the case of isothermal outflow the mass loss rate
is {\it anti-correlated} with the mass accretion rate in the 
Keplerian component. This was due to the fact that
when the accretion rate in the Keplerian disk is low, the inner, advective
region remains hotter and therefore drives more mass loss.

This phenomenon has an interesting consequence which we explore in this
{\it Rapid Communication}. As more fraction of matter is expelled, it becomes 
easier for the soft photons from the Keplerian disk surrounding the advective 
region to cool this region due to Comptonization (see, Chakrabarti \& Titarchuk 
[13] and Chakrabarti [14] for details). In other words, the presence of winds
would {\it soften} the spectra of the power-law component for the {\it same}
multicolour blackbody component. Such an observation
would point to profuse mass loss from the hot advective region.

The model we use here is the two component accretion flow which has 
the centrifugal pressure supported boundary layer [CENBOL]. We assume
a weakly viscous flow of constant specific angular momentum $\lambda=1.75$
(in units of $2GM/c$) for the sake of concreteness. The Keplerian
component close to the equatorial plane has a low accretion rate
(${\dot M}_{in} \sim 0.05-0.3$ in units of the Eddington rate) 
and the sub-Keplerian halo surrounding it has a higher rate 
(${\dot M}_h \sim 1$ in units of Eddington rate). Before the 
accreting matter hits the inner advective region, both the rates are constant,
but as Das \& Chakrabarti [12] has shown, winds, produced from CENBOL
will deplete the disk matter at the rate determined by the temperature of the
CENBOL, when other parameters, such as the specific angular momentum and specific
energy are kept fixed. In particular, it was shown that in some cases, high entropy
matter from CENBOL may be completely bounced back as winds and produce 
quiescence states or black hole candidates with very low luminosity such as
the Sgr $A^*$ in our Galactic Center and V404 Cyg.

In Fig. 1 we schematically draw a global picture of the 
accretion and outflow around the black hole. As matter accretes on a black 
hole, Keplerian flows tends to stay at the higher viscosity region,
namely on the equatorial plane. The low viscosity region away from the
plane has positive specific energy and forms an advective inflow
which may or may not form a shock wave closer to the centrifugal barrier.
Keplerian disks emit soft blackbody photons which are intercepted and
reprocessed in CENBOL and its surrounding sub-Keplerian region. 
Radiation out of this region through Comptonization and bremsstrahlung 
processes come out carrying the information about the optical
depth and temperature of this region. Thus a proper prediction of the
power-law spectra in hard X-rays would be valuable to understand
black hole accretion and wind processes.

Figures 2 and 3 show the outcome of our calculation of the spectra for 
different accretion rate of the Keplerian component ${\dot M}_{in}$. The mass
of the central black hole is chosen to be $M=10M_\odot$. The 
location of the CENBOL is assumed to be at $r=10r_g$ (where $r_g$ is the
Schwarzschild radius), a typical location for the sub-Keplerian flow
of average specific angular momentum $\lambda=1.75$ and specific 
energy ${\cal E}=0.003$. Following [12], we first compute the mass outflow rate
from the advective region. The long dashed curve in Fig. 2 shows the 
variation of the percentage of mass loss (vertical axis on the right)
as a function of the inflow accretion rate. The dotted curve and the
solid curve denote the variation  of the energy spectral index $\alpha$
($F_\nu \propto \nu^{-\alpha}$) with and without winds taken into account.
Note that  the spectra is overall softened ($\alpha$ increased)
when winds are present. For higher Keplerian rates, the mass loss through winds is 
negligible and therefore there is virtually no change in the spectral
index. For lower inflow rates, on the other hand, mass loss rate is more than twenty
percent. It is easier to Comptonize the depleted matter by the same number
of incoming soft photons and therefore the spectra is softened.

In Fig. 3, we show the resulting spectral change. As in Fig. 2, solid
curves represent solutions without winds and the dotted curves
represent solutions with winds. Solid curves are drawn for ${\dot M}_{in}=0.3$ 
(uppermost at the bump), $0.15$ (middle at the bump) and $0.07$ (lowermost at the bump)
respectively. For ${\dot M}_{in}=0.3$ both curves are identical.
Note the crossing of the solid curves at around $10^{18.6}Hz$
($15$ keV) when winds are absent. This is regularly observed in black hole
candidates. If this is shifted to higher energies, the presence of 
winds may be indicated.

Strong winds are suspected to be present in Sgr $A^*$ at our Galactic Center
(see, Genzel et al [15] for a review, and Eckart \& Genzel [16] ). Chakrabarti
[1] suggested that the inflow could be of almost constant energy transonic
flow, so that the emission is inefficient. However, from global inflow-outflow
solutions [GIOS], Das \& Chakrabarti [12] showed that when the inflow rate
itself is low ( as is the case for Sgr A$^*$; $\sim 10^{-3}$ to $10^{-4} {\dot M}_{Eddington}$) 
the mass outflow rate is very high, almost to the point of evacuating the disk.
This prompted them to speculate that spectral properties of our
Galactic Center could be explained by inclusion of winds. This 
will be done in near future. Not only our Galactic Center, the consideration
should be valid for all the black hole candidates (e.g., V404 Cyg) which are seen
in quiescence. 

Earlier,  Chakrabarti \& Titarchuk [13] suggested that the iron K$_\alpha$ line as well 
as the so called `reflection component' could be due to outflows
off the advective region. Combined with the present work, we may conclude 
that simultaneous enhancement of the `reflection component' and/or 
iron K$_\alpha$ line intensity with the softening of the spectra 
in hard X-rays would be a sure signature of the presence of 
signficant winds in the advective region of the disk.

\newpage

\centerline {Reference}

\bibliographystyle{plain}
{}
\newpage 
\centerline{Figure Captions}

Fig. 1: Cartoon diagram of a very general inflow-outflow configuration 
of non-magnetized matter around a compact object. Keplerian and sub-Keplerian 
matter accretes along the equatorial plane. Centrifugally and thermally driven 
outflows preferentially emerge between the centrifugal barrier 
and the funnel wall. $X_s$, $X_{K1}$ and $X_{K2}$ are the boundaries of the (possible) shock,
between Keplerian and sub-Keplerian flow of high viscosity component, and between Keplerian
and sub-Keplerian flow of very weak viscosity  components respectively.

Fig. 2: Variation of the percentage of mass loss (long dashed curve and right axis) 
and the energy spectral index $\alpha$ ($F_\nu \propto \alpha^{-\alpha}$) (solid and dotted curves and
left axis) with the accretion rate (in units of Eddington rate) of the Keplerian component.
Solid curve is drawn when winds are neglected from the advective region and dotted curve includes
effect of winds. Overall spectra is softened when the inflow rate is reduced.

Fig. 3: Spectra of emitted radiation from the accretion disk with (dotted) and without (solid)
effects of winds.  Hard X-ray component is softened while keeping soft  X-ray bump unchanged.

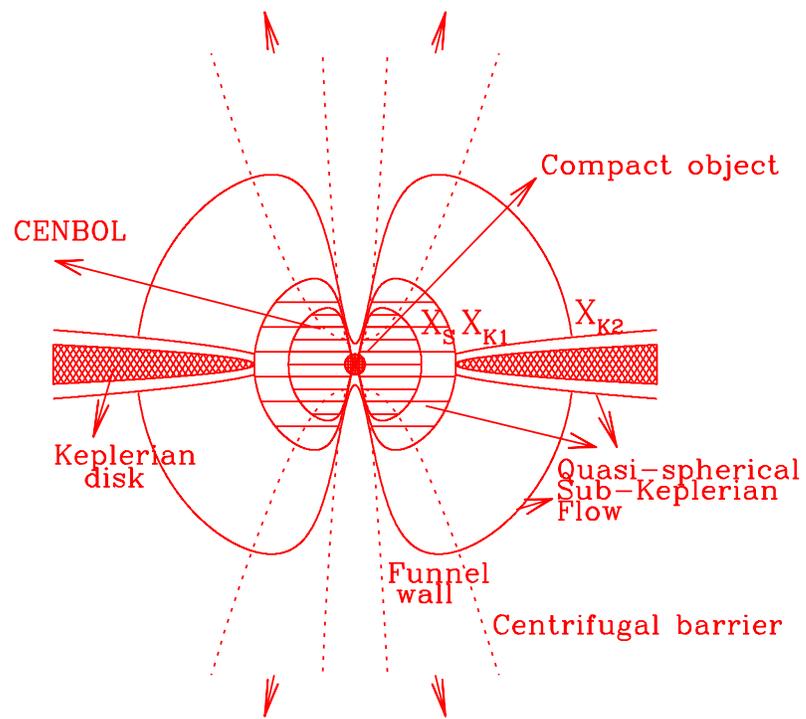
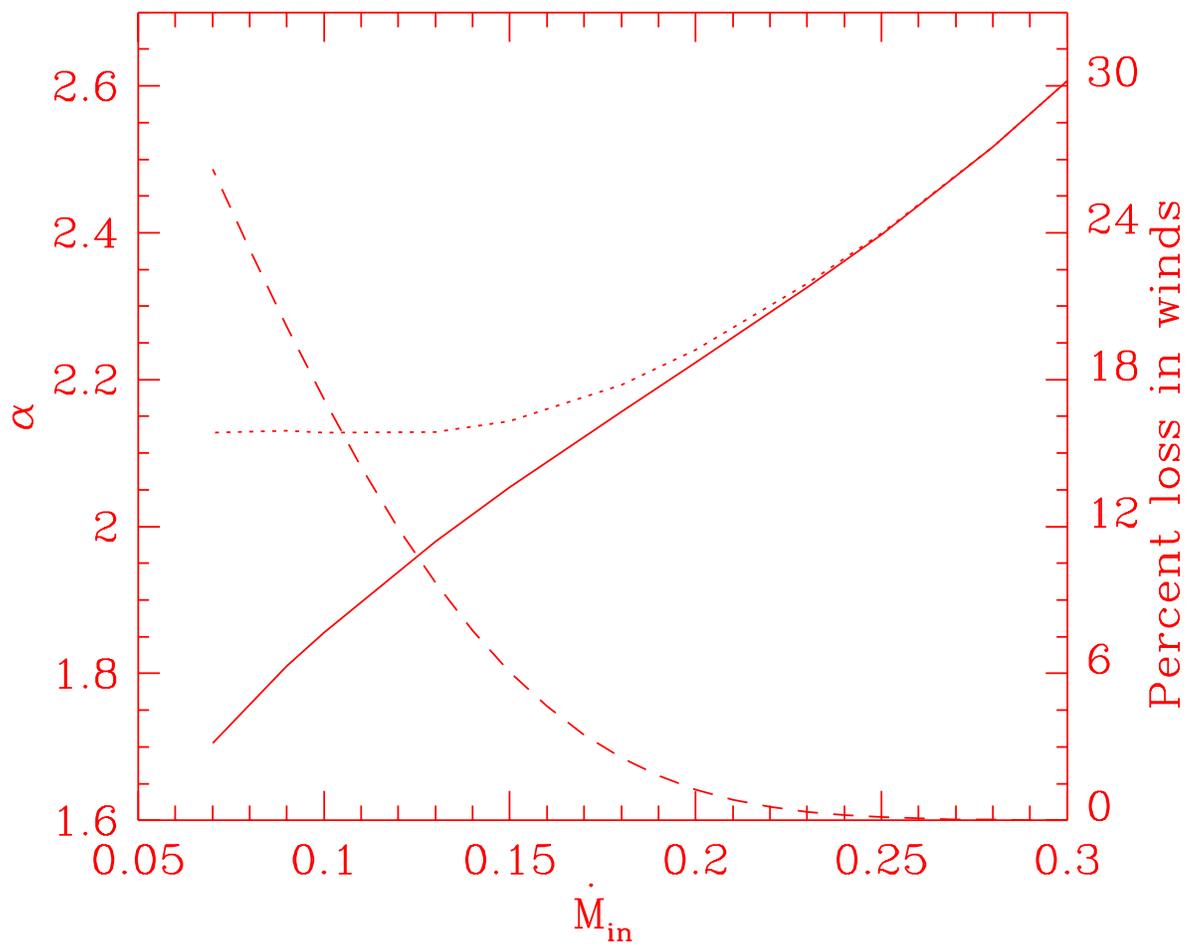
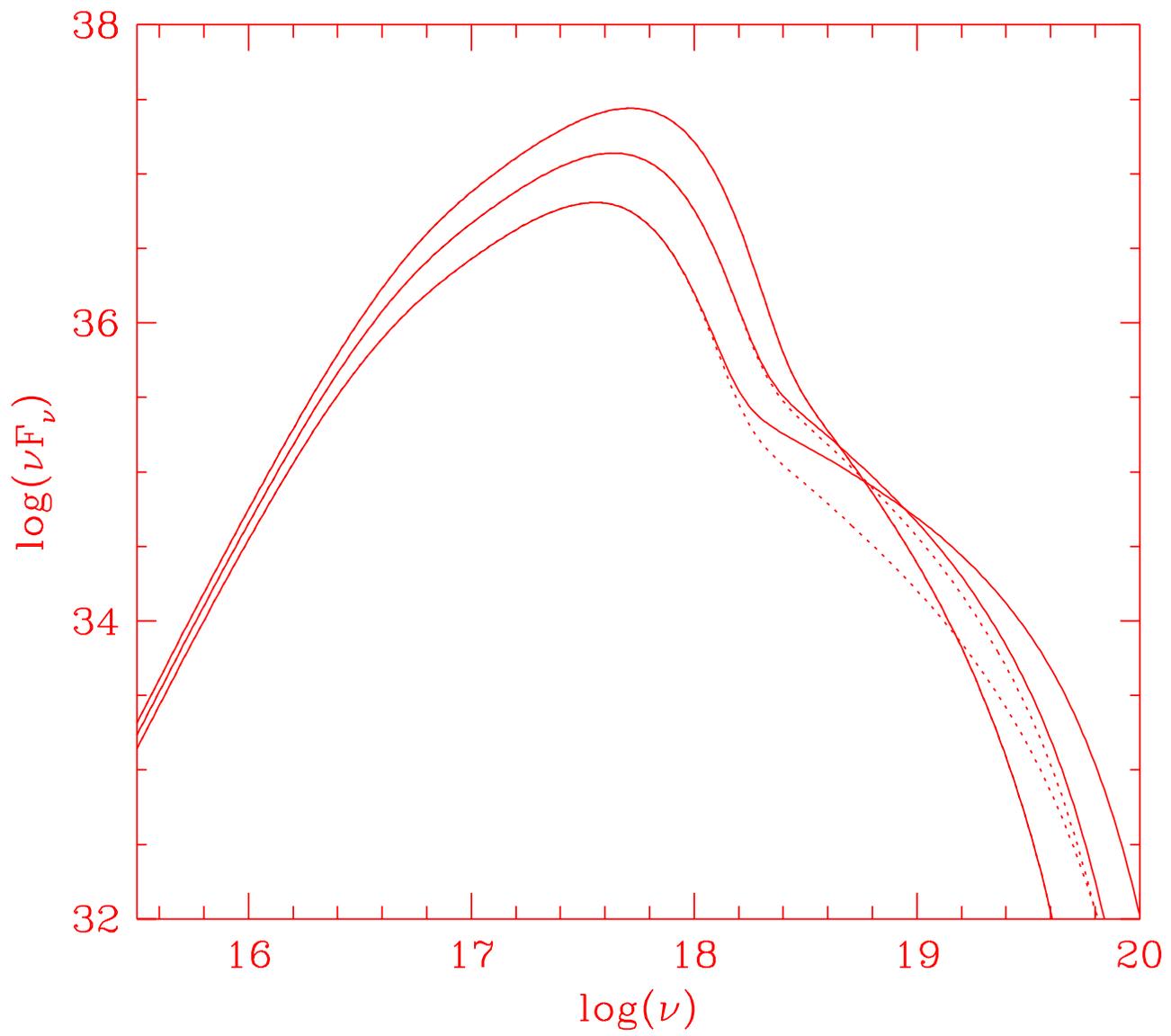
\end{document}